\documentclass[12pt, hidelinks, oneside]{hdsr}

\usepackage{pbox}
\usepackage{centernot}
\newtheorem{remark}{Remark}
\newtheorem{assumption}{Assumption}

\newtheorem{theorem}{Theorem}[section]

\newcommand{\sign}{\text{sign}}

\newcommand{\cC}{{\mathcal{C}}}
\newcommand{\cL}{{\mathcal{L}}}
\newcommand{\cU}{{\mathcal{U}}}
\newcommand{\cD}{{\mathcal{D}}}

\newcommand{\indep}{\perp \!\!\! \perp}

\usepackage{tikz}
\usetikzlibrary{arrows.meta,
                chains,
                positioning,
                shapes.geometric
                }
          \usetikzlibrary{calc,trees,positioning,chains,decorations.pathreplacing,decorations.pathmorphing,shapes,matrix,shapes.symbols,plotmarks,decorations.markings,shadows,automata,backgrounds,petri}

\usepackage{fancyhdr}
\begin{document}

\newgeometry{bottom=1.5in}

\volumeheader{3}{3}{10.1162/99608f92.d07b8d16}

\begin{center}

  \title{Individualized Decision-Making Under Partial Identification: Three Perspectives, Two Optimality Results, and One Paradox}
  \maketitle

  \thispagestyle{empty}


  \begin{tabular}{cc}
    Yifan Cui\upstairs{*}
   \\[0.25ex]
   {
 National University of Singapore} \\
  \end{tabular}

  \emails{\vspace{-1cm}
    \upstairs{*}cuiy@nus.edu.sg
    }

\begin{abstract}
Unmeasured confounding is a threat to causal inference and gives rise to biased estimates.
In this article, we consider the problem of individualized decision-making under partial identification.
Firstly, we argue that when faced with unmeasured confounding,
one should pursue individualized decision-making using partial identification in a comprehensive manner.
We establish a formal link between individualized decision-making under partial identification and classical decision theory by considering a lower bound perspective of value/utility function.
Secondly,
building on this unified framework, we provide a novel minimax solution (i.e., a rule that minimizes the maximum regret for so-called opportunists) for individualized decision-making/policy assignment.
Lastly, we provide an interesting paradox drawing on novel connections between two challenging domains, that is, individualized decision-making and unmeasured confounding.
Although motivated by instrumental variable bounds, we emphasize that the general framework proposed in this article would in principle apply for a rich set of bounds that might be available under partial identification.
\end{abstract}
\end{center}

\vspace*{0.15in}
\hspace{10pt}

\copyrightnotice

\section*{Media Summary}

In the era of big data, observational studies are a treasure for both association analysis and causal inference, with the potential to improve decision-making.
Depending on the set of assumptions one is willing to make, one might achieve either point, sign, or partial identification of causal effects.
In particular, under partial identification, it might be inevitable to make suboptimal decisions.
Policymakers caring about decision-making would face the following important question: What are optimal strategies corresponding to different risk preferences?\\
\indent In this article, the author offers a unified framework that generalizes several decision-making strategies in the literature. Building on this unified framework, the author also provides a novel minimax solution (i.e., a rule that minimizes the maximum regret for so-called opportunists) for individualized decision-making and policy assignment.


\newpage

\section{The power of storytelling: different views might lead to different decisions}
\label{sec1}

Suppose one is playing a two-armed slot machine.
The rewards $R_{-1}$ and $R_{1}$ are the payoffs for hitting the jackpot of each arm, respectively.
For simplicity, let us assume that both arms always give positive rewards ($R_{-1},R_{1}>0$), that is, one is guaranteed not to lose and therefore would not refrain from playing this game.
However, due to some uncertainty, one does not have prior knowledge of the exact values of $R_{-1}$ and $R_1$. Fortunately, suppose there is a magic instrument, which can help one to identify the range of rewards.

\begin{figure}[H]
\centering
\begin{tikzpicture}
\begin{scope}
    \draw [thick](2,0.2)--(2,1.6);
    \draw [thick](4,0.6)--(4,1);
    \draw [thick](1.75,0.2)--node[right=2mm]{$0.1$}(2.25,0.2);
    \draw [thick](1.75,1.6)--node[right=2mm]{$0.8$}(2.25,1.6);
    \draw [thick](3.75,0.6)--node[right=2mm]{$0.3$}(4.25,0.6);
    \draw [thick](3.75,1)--node[right=2mm]{$0.5$}(4.25,1);
    \draw [dashed](0.5,0)--node[xshift=2.5cm,yshift=-3mm]{$0$}(5.5,0);
        \node[rectangle, thick, draw=black!10,fill=black!10,minimum size=10mm, minimum height=7mm, xshift=2cm, yshift=-0.5cm]{$R_{-1}$};
    \node[rectangle, thick, draw=black!10,fill=black!10,minimum size=10mm, minimum height=7mm, xshift=4cm, yshift=-0.5cm]{$R_1$};
\end{scope}
\begin{scope}[xshift=-3cm,yshift=1cm]
        \draw [thick,blue](0,-1)--(0,0.8);
    \draw [thick,blue](-0.25,-1)--node[right=2mm,blue]{$-0.5$}(0.25,-1);
    \draw [thick,blue](-0.25,0.8)--node[right=2mm,blue]{$0.4$}(0.25,0.8);
    \draw [dashed](-2.5,0)--node[xshift=2.5cm,yshift=-3mm]{$0$}(2.5,0);
        \node[rectangle, thick, draw=black!10,fill=black!10,minimum size=10mm, minimum height=7mm, xshift=0cm, yshift=-1.5cm]{$R_1-R_{-1}$};
\end{scope}
 \begin{pgfonlayer}{background}
    \filldraw [line width=4mm,black!10,xshift=-3cm,yshift=2cm]
      (-2,-3)  rectangle (2,1);
    \filldraw [line width=4mm,black!10]
      (1,-1) rectangle (5,3);
  \end{pgfonlayer}
\end{tikzpicture}
\caption{A toy example on slot machines. The left panel: the possible range of $R_1-R_{-1}$; the right panel: the possible ranges of $R_{-1}$ and $R_1$, respectively. \label{fig:0}}
\end{figure}
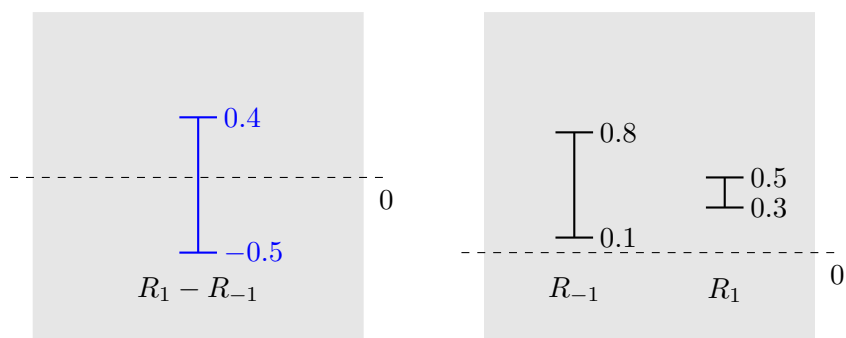

By only providing one with the left panel of Figure~\ref{fig:0}, that is, the range of $R_1-R_{-1}$, most people might opt to pull arm $-1$.
But wait a minute... where am I, and why am I looking at the left panel without knowing the real payoffs?
After looking at the right panel, the decision might be changed depending on a person's risk preference.

\tikzstyle{txt} = [rectangle, thick, draw=white,fill=white,minimum size=10mm, minimum height=7mm]

Is there such an instrument in real life? The answer is in the affirmative. One such instrument is a so-called instrumental variable (IV).
In statistics and related disciplines, an IV method is used to estimate causal relationships when randomized experiments are not feasible or when there is noncompliance in a randomized experiment.  Intuitively, a valid IV induces changes in the explanatory variable but otherwise has no direct effect on the dependent variable, allowing one to uncover the causal effect of the explanatory variable on the dependent variable.
Under certain IV models, one can obtain bounds for counterfactual means.
So how would one pursue decision-making when faced with partial identification?
The rest of the article offers a comprehensive view of individualized decision-making under partial identification as well as several novel solutions to various decision- and policy-making strategies.

\subsection{Introduction}

An optimal decision rule provides a personalized action/treatment strategy for each participant in the population based on one's individual characteristics.
A prevailing strand of work has been devoted to estimating optimal decision rules \citep[and many others]{murphy2001marginal,murphy2003optimal,robins2004optimal,qian2011performance,zhao2012estimating,zhang2012estimating,wager2021policy}; we refer to \citet{chakraborty2013statistical,kosorok2019review,tsiatis2019dynamic} for an up-to-date literature review on this topic.

Recently,
there has been a fast-growing literature on estimating individualized decision rules based on observational studies subject to potential unmeasured confounding \citep{kallus2018confounding,yadlowsky2018bounds,kallusinterval,cui2020,qiu2020,pz,zp,han2018identification,han2020,han2019optimal,cuirejoinder,qiu2021,cui2020necessary}.
In particular, \citet{cui2020} pointed out that one could identify treatment regimes that maximize lower bounds of the value function when one has only partial identification through an IV.
\citet{pz} further proposed an IV-optimality criterion to learn an optimal treatment regime, which essentially recommends the treatment for patients for whom the estimated conditional average treatment effect bound covers zero based on the length of the bounds, that is, based on the left panel of Figure~\ref{fig:0}. See more details in \citet{cui2020,zp,cuirejoinder}.

In this article, we provide a comprehensive view of individualized decision-making under partial identification
through maximizing the lower bounds of the value function. This new perspective unifies various classical decision-making strategies in classical decision theory. Building on this unified framework, we also provide a novel minimax solution (for so-called opportunists who are unwilling to lose) for individualized decision-making and policy assignment.
In addition, we point out that there is a mismatch between different optimality results, that is, an `optimal' rule that attains one criterion does not necessarily attain the other.
Such mismatch is a distinctive feature of individualized decision-making under partial identification, and therefore makes the concept of universal optimality for decision-making under uncertainty ill-defined.
Lastly, we provide a paradox to illustrate that a non-individualized decision can conceivably lead to an outcome superior to an individualized decision under partial identification.
The provided paradox also sheds light on using IV bounds as sanity check or policy improvement.

To conclude this section, we briefly introduce notation used throughout the article. Let $Y$ denote the outcome of interest and $A \in \{-1,1\}$ be a binary action/treatment indicator.
Throughout it is assumed that larger values of $Y$ are more desirable.
Suppose that $U$ is an unmeasured confounder of the effect of $A$ on $Y$.  Suppose also that one has observed a pretreatment binary IV $Z \in \{-1,1\}$.
Let $X$ denote a set of fully observed pre-IV covariates.
Throughout, we assume the complete data are independent and identically distributed realizations of $(Y, X, A, Z, U)$; thus the observed data are $(Y,X,A,Z)$.

\section{A brief review of optimal decision rules with no unmeasured confounding}

An individualized decision rule is a mapping from the covariate space to the action
space $\{-1, 1\}$.
Suppose $Y_a$  is a person's potential outcome under an intervention that sets $A$ to value $a$, $Y_{\cD(X)}$ is the potential outcome under a hypothetical intervention that assigns $A$ according to the rule $\cD$, that is,
$Y_{\cD(X)} \equiv Y_{1}I\{\cD(X)=1\}+Y_{-1}I\{\cD(X)=-1\}$,
$E[Y_{\cD(X)}]$ is the value function \citep{qian2011performance}, and $I\{\cdot\}$ is the indicator function.
Throughout the article, we make the following standard consistency and positivity assumptions:
(1) For a given regime $\cD$, $Y = Y_{\cD(X)}$ when $A = \cD(X)$ almost surely. That is, a person’s observed outcome matches his/her potential outcome under a given decision rule when the realized action matches his/her potential assignment under the rule; (2) We assume that $\Pr(A = a|X) > 0$ for $a = \pm 1$ almost surely. That is, for any observed covariates~$X$, a person has an opportunity to take either action.

We wish to identify an optimal decision rule $\cD^*$ that admits the following representation, that is,
\begin{align}
\cD^*(X) = \sign\{ E(Y_1-Y_{-1}|X)>0 \}
~\text{or}~
\cD^* = \arg\max_{\cD} E[Y_{\cD(X)}].
\label{eq:opt2}
\end{align}
A significant amount of work has been devoted to estimating optimal decision rules relying on the following unconfoundedness assumption:
\begin{assumption}\emph{(Unconfoundedness)}
$Y_a \indep A| X$ for $a=\pm 1$.
\label{asm:unconfoundedness}
\end{assumption}
The assumption essentially rules out the existence of an unmeasured factor $U$ that confounds the effect of $A$ on $Y$ upon conditioning on $X$. It is straightforward to verify that
under Assumption~\ref{asm:unconfoundedness}, one can identify the value function $E[Y_{\cD(X)}]$ for a given decision rule $\cD$.
Furthermore, the optimal decision rule in Equation~\eqref{eq:opt2} is identified from the observed data
\begin{align*}
\cD^*(X) = \sign\{ \cC(X)>0 \},
\end{align*}
where $\cC(X)=E(Y|X,A=1) - E(Y|X,A=-1)=E(Y_1-Y_{-1}|X)$ denotes the conditional average treatment effect (CATE).
As established by \citet{qian2011performance}, learning optimal decision rules under Assumption~\ref{asm:unconfoundedness} can be formulated as
\begin{align*}
\cD^*=\arg\max_{\cD} E\left[\frac{I\{\cD(X)=A\}Y}{\Pr(A|X)}\right],
\end{align*}
where $\Pr(A|X)$ is the probability of taking $A$ given $X$.
\citet{zhang2012robust} proposed to directly maximize the value function over a parametrized set of functions.
Rather than maximizing the above value function,
\citet{zhao2012estimating,zhang2012estimating,rubin2012statistical} transformed the above problem into a weighted classification problem,
\begin{align*}
 \arg\min_\cD E \{|\cC(X)| I[\sign\{\cC(X)>0\} \neq \cD(X)]\}.
\end{align*}
The ensuing classification approach was shown to have appealing robustness properties, particularly in a randomized study where no model assumption on $Y$ is needed.

\section{Instrumental variable with partial identification}
In this section, instead of relying on Assumption~\ref{asm:unconfoundedness}, we allow for unmeasured confounding, which might cause biased estimates of optimal decision rules.
Let $Y_{z,a}$ denote the potential outcome had, possibly contrary to fact, a person's IV and treatment value been set to $z$ and $a$, respectively.
 Suppose that the following assumption holds:
\begin{assumption}\emph{(Latent unconfoundedness)}
$Y_{z,a} \indep (Z, A)|X, U$ for $z,a = \pm 1$.
\label{asm:unconfoundedness2}
\end{assumption}
This assumption essentially states that together $U$ and $X$ would in principle suffice to account for any confounding bias.
Because $U$ is not observed, we propose to account for it when a valid IV $Z$ is available that satisfies the following standard IV assumptions \citep{cui2020}:

\begin{assumption}\emph{(IV relevance)} $Z \centernot{\indep} A|X$.
\label{IV Relevance}
\end{assumption}

\begin{assumption}\emph{(Exclusion restriction)} $Y_{z,a}=Y_a$ for $z,a=\pm 1$ almost surely.
\label{Exclusion Restriction}
\end{assumption}

\begin{assumption}\emph{(IV independence)} $Z \indep  U |X$.
\label{IV Independence}
\end{assumption}

\begin{assumption}\emph{(IV positivity)} $0<\Pr\left(  Z=1|X\right)<1$ almost surely.
\label{asm:IV positivity}
\end{assumption}

Assumptions~\ref{IV Relevance}-\ref{IV Independence} are well-known IV conditions, while Assumption~\ref{asm:IV positivity} is needed for nonparametric identification \citep{imbens1994,angrist1996,greenland2000,hernan2006epi}.
Assumption~\ref{IV Relevance} requires that the IV is associated with the treatment conditional on $X$. Note that Assumption~\ref{IV Relevance} does not rule out confounding of the $Z$-$A$ association by an unmeasured factor, however, if present, such factor must be independent of $U$. Assumption~\ref{Exclusion Restriction} states that there can be no direct causal effect of $Z$ on $Y$ not mediated by $A$. Assumption~\ref{IV Independence} states that the direct causal effect of $Z$ on $Y$ would be identified conditional on $X$ if one were to intervene on $A=a$.
Figure~\ref{fig:1} provides a graphical representation of
Assumptions \ref{Exclusion Restriction} and \ref{IV Independence}.

\tikzstyle{var} = [circle,  very thick,draw=black,fill=white,minimum size=10mm]
\begin{figure}[H]
\centering
\begin{tikzpicture}[>=latex]
    \node (A) [var, xshift=0cm,yshift=0cm]{$\boldsymbol{A}$};
    \node (U) [var, xshift=1.5cm,yshift=-1.5cm]{$\boldsymbol{U}$};
    \node (Y) [var, xshift=3cm,yshift=0cm]{$\boldsymbol{Y}$};
    \node (X) [var, xshift=1.5cm,yshift=1.5cm]{$\boldsymbol{X}$};
    \node (Z) [var, xshift=-3cm,yshift=0cm]{$\boldsymbol{Z}$}
    edge[<->, bend right=45, very thick](A)
    ;
    \draw[->, very thick] (X) -- (Z);
    \draw[->, very thick] (X) -- (A);
    \draw[->, very thick] (X) -- (Y);
    \draw[->, very thick] (Z) -- (A);
    \draw[->, very thick] (A) -- (Y);
    \draw[->, very thick] (U) -- (A);
    \draw[->, very thick] (U) -- (Y);
\end{tikzpicture}
\caption{A causal graph with unmeasured confounding. The bi-directed arrow between $Z$ and $A$ indicates the possibility that there may be unmeasured common causes confounding their association. \label{fig:1}}
\end{figure}
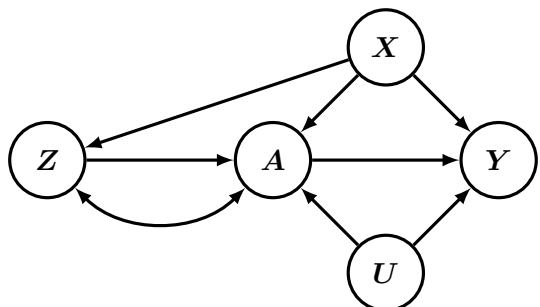
While Assumptions~\ref{IV Relevance}-\ref{asm:IV positivity} together do not suffice for point identification of the counterfactual mean and average treatment effect, a valid IV, even under minimal four assumptions, can partially identify the counterfactual mean and average treatment effect, that is,
lower and upper bounds might be formed.
Let $\mathcal{L}_{-1}\left( X\right) $, $\mathcal{U}_{-1}\left( X\right) $, $\mathcal{L}_{1}\left( X\right) $, $\mathcal{U}_{1}\left( X\right) $ denote lower and upper bounds for $E\left( Y_{-1}|X\right) $ and $E\left( Y_{1}|X\right) $; hereafter, we consider lower and upper bounds for  $E\left( Y_{1}-Y_{-1}|X\right)$ of form
$\mathcal{L}\left( X\right)=\cL_1(X)-\cU_{-1}(X)$ and $\mathcal{U}\left( X\right)=\cU_1(X)-\cL_{-1}(X)$, respectively; sharp bounds for $E\left( Y_{1}-Y_{-1}|X\right)$ in certain prominent IV models have been shown to take such a form, see for instance Robins-Manski bound \citep{robins1989, manski1990nonparametric},
Balke-Pearl bound \citep{Balke1997}, Manski-Pepper bound under a monotone IV assumption \citep{manski2000monotone}
and many others.
Here, we
consider the following conditional Balke-Pearl bounds
\citep{cui2020} for a binary outcome as our running example. Let $p_{y,a,z,x}$ denote $\Pr(Y = y, A = a|Z = z, X = x)$, and
\begin{eqnarray*}
\mathcal{L}_{-1}\left( x\right) =& \max
\left \{
  \begin{tabular}{c}
  $p_{1,-1,1,x}$ \\
  $p_{1,-1,-1,x}$ \\
  $p_{1,-1,-1,x} + p_{1,1,-1,x} - p_{-1,-1,1,x} - p_{1,1,1,x} $ \\
  $p_{-1,1,-1,x} + p_{1,-1,-1,x} - p_{-1,-1,1,x} - p_{-1,1,1,x} $ \\
  \end{tabular}
\right \}, \\
\mathcal{U}_{-1}\left( x \right) =& \min
\left \{
  \begin{tabular}{c}
  $1 - p_{-1,-1,1,x}$ \\
  $1- p_{-1,-1,-1,x}$ \\
  $p_{-1,1,-1,x} + p_{1,-1,-1,x} + p_{1,-1,1,x} + p_{1,1,1,x} $ \\
  $p_{1,-1,-1,x} + p_{1,1,-1,x} + p_{-1,1,1,x} + p_{1,-1,1,x} $ \\
  \end{tabular}
\right \}, \\
\mathcal{L}_{1}\left( x \right) =& \max
\left \{
  \begin{tabular}{c}
  $p_{1,1,-1,x}$ \\
  $p_{1,1,1,x}$ \\
  $-p_{-1,-1,-1,x} - p_{-1,1,-1,x} + p_{-1,-1,1,x} + p_{1,1,1,x} $ \\
  $-p_{-1,1,-1,x} - p_{1,-1,-1,x} + p_{1,-1,1,x} + p_{1,1,1,x} $ \\
  \end{tabular}
\right \}, \\
\mathcal{U}_{1}\left( x \right) =& \min
\left \{
  \begin{tabular}{c}
  $1 - p_{-1,1,1,x}$ \\
  $1- p_{-1,1,-1,x}$ \\
  $p_{-1,-1,-1,x} + p_{1,1,-1,x} + p_{1,-1,1,x} + p_{1,1,1,x} $ \\
  $p_{1,-1,-1,x} + p_{1,1,-1,x} + p_{-1,-1,1,x} + p_{1,1,1,x} $ \\
  \end{tabular}
\right \}.
\end{eqnarray*}
Additionally, one could proceed with other partial identification assumptions and corresponding bounds. We refer to references cited in \citet{Balke1997} and a review paper by \citet{swanson2018partial} for alternative bounds.

We conclude this section by providing multiple settings in real life where an IV is available but Assumption~\ref{asm:unconfoundedness} is not likely to hold:
1) In a double-blind placebo-randomized trial in which participants are subject to noncompliance, the treatment assignment is a valid IV; 2) Another classical example is that in sequential, multiple assignment, randomized trials (SMARTs) in which patients are subject to noncompliance, the adaptive intervention is a valid IV.
We note that the later proposed randomized minimax solution in Section~\ref{sec:minimax_game} offers a promising strategy for this setting; 3) In social studies, a classical example is estimating the causal effect of education on earnings. Residential proximity to a college is a valid IV. We will further elaborate the third example in the next section.

\section{A real-world example} \label{sec:real}
In this section, we first consider a real-world application on the effect of
education on earnings using data from the National Longitudinal Study of Young Men \citep{card1993using, tan2006regression,okui2012doubly,wang2017,wang2018},
which consist of 5,525 participants aged between 14 and 24 in 1966. Among them, 3,010 provided valid education and wage responses in the 1976 follow-up.
Following \citet{tan2006regression,wang2018}, we consider education beyond high school as a binary action/treatment (i.e., $A$).
A practically relevant question is the following: Which students would be better off starting college to maximize their earnings?

In this study, there might be unmeasured confounders even after adjusting for  observed
covariates, for example, unobserved preferences for education levels might be an unmeasured confounder that is likely to be associated with both education and wage.
We follow \citet{card1993using,wang2017,wang2018} and use presence of a nearby four-year college as an instrument (i.e., $Z$).
In this data set, 2,053 (68.2\%) lived close to a four-year college, and 1,521 (50.5\%) had education
beyond high school. To illustrate the IV bounds with binary outcomes, we follow \citet{wang2017,wang2018} to dichotomize the outcome wage (i.e., $Y$) at its median, that is 5.375 dollars per hour. While we only use this as an illustrating example, we note that dichotomizing earnings might affect decision-making, and therefore in practice one might conduct a sensitivity analysis around the choice of cut-off. Following \citet{wang2018}, we adjust for age, race, father and mother’s education levels, indicators for residence in the south and
a metropolitan area and IQ scores (i.e., $X$), all measured in 1966. Among them, race, parents’ education levels, and
residence are included as they may affect both the IV and outcome; age is included as it is likely to modify the effect of education on earnings; and IQ scores, as a measure of underlying ability, are included
as they may modify both the effect of proximity to college on education, and the effect of education on earnings.

\begin{figure}[ht]
    \centering
    \includegraphics[width=5.2cm, height=5.2cm]{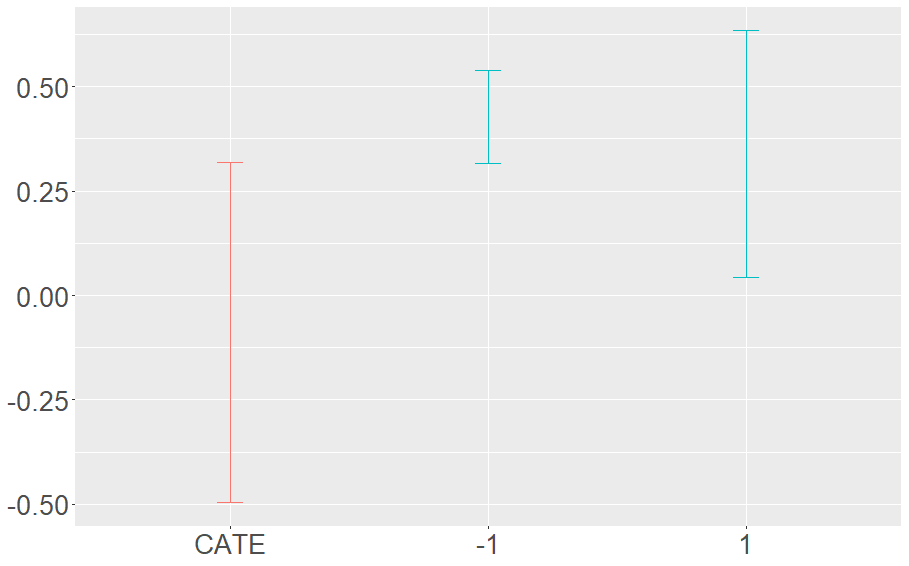} \hspace{0.7cm}
      \includegraphics[width=5.2cm, height=5.2cm]{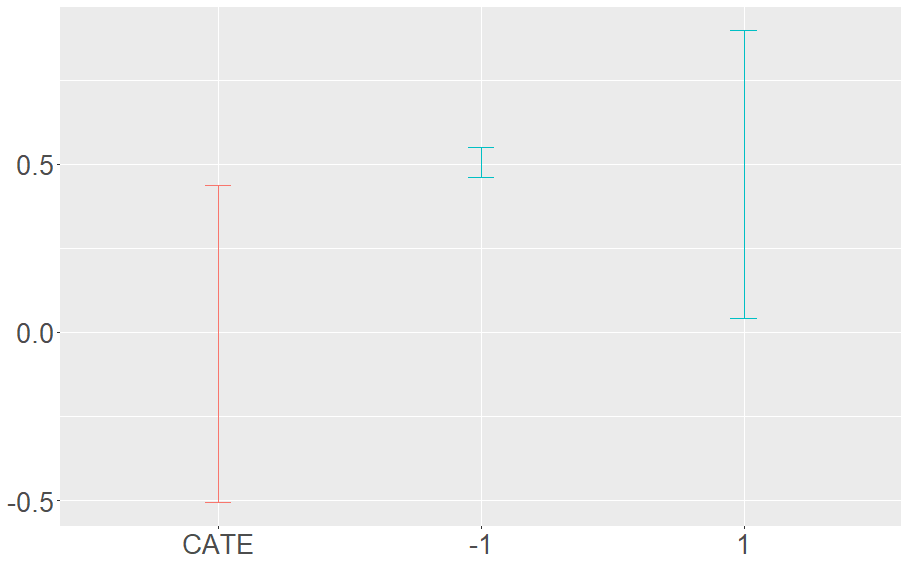} \\
    \caption{IV CATE and counterfactual mean bounds for two subjects with IQ scores 84.00 and 102.45, where $A=1$ and $-1$ refer to education beyond high school or not, respectively.}
    \label{fig:2}
\end{figure}

We use random forests to estimate the probability of $p_{y,a,z,x}$ \citep[with default tuning parameters in][]{RF} and then construct estimates of Balke-Pearl bounds $\mathcal{L}_{-1}\left( X\right) $, $\mathcal{U}_{-1}\left( X\right) $, $\mathcal{L}_{1}\left( X\right) $, $\mathcal{U}_{1}\left( X\right) $,
$\mathcal{L}\left( X\right) $, $\mathcal{U}\left( X\right) $.
To streamline our presentation, we consider the subset of individuals of age 15, parents' education level 11 years, non-Black, and residence in a non-south and metropolitan area. Their IV CATE and counterfactual mean bounds $\cL(X)$, $\cU(X)$, $\cL_{-1}(X)$, $\cU_{-1}(X)$, $\cL_{1}(X)$, $\cU_{1}(X)$ are presented in Figure~\ref{fig:2}.

The shape of IV bounds looks similar to the slot machine example of Figure~\ref{fig:0} given at the beginning of the article. When faced with uncertainty, what are different decision-making strategies?
In the next section, we provide a new perspective of viewing optimal decision-making under partial identification beyond just looking at contrast or value function.
Except for the real-world example, for pedagogical purposes, we focus on the population level of IV bounds instead of their empirical analogs throughout.

\section{The lower bound perspective: A unified criterion} \label{sec:5}

In Section~\ref{sec:5.1}, we link the lower bound framework to well established decision theory from an investigator's perspective.
In Section~\ref{sec:5.2}, we extend our framework to take into account individual preferences of participants.
In Section~\ref{sec:minimax_game}, we provide a formal solution to achieve a minimax regret goal by leveraging a randomization scheme. In Section~\ref{sec:no_optimality}, we reveal a mismatch between deterministic/randomized minimax regret and maximin utility, and conclude that there is no universal concept of optimality for decision-making under partial identification.

\subsection{A generalization of classical decision theory} \label{sec:5.1}
In this section, we establish a formal link between individualized decision-making under partial identification and classical decision theory.
The set of rules $\cD(w(x),x)$ which maximize the following lower bounds of $E[Y_{\cD(X)}]$,
\begin{align*}
&\Big\{E_X \{ [1-w(X)] [\mathcal{L}\left( X\right)
 I\left\{ \mathcal{D}(X)=1\right\} +\mathcal{L}_{-1}\left( X\right)]
 + w(X) [ -\mathcal{U}\left( X\right) I\left\{ \mathcal{D}(x)=-1\right\} +\mathcal{L}_{1}\left( X\right)]\}:\\&
 ~~~~ \text{where}~w(x)~\text{can depend on $\cD(x)$},~0\leq w(x) \leq 1,~\text{for any}~ x\Big\},
\end{align*}
 is denoted by $\cD^{opt}$.
The derivation of lower bounds of $E[Y_{\cD(X)}]$ is provided in the Appendix.
 Hereinafter, we refer to reasoning decision-making strategy from $\cD^{opt}$ as the lower bound criterion,
where, as can be seen later, $w(x)$ reflects the investigator's preferences.

In Table~\ref{table:1}, we provide examples of decision-making criteria that have previously appeared in classical decision theory and we connect each such criterion to a corresponding $w(x)$.
Hereafter, for a rule $\cD$, we formally define utility as value function $E[Y_{\cD(X)}]$ and regret as $E[Y_{\cD^*(X)}] - E[Y_{\cD(X)}]$.
We give the formal definition of each rule in Table~ \ref{table:1} except that the mixed strategy is deferred to Section~\ref{sec:minimax_game}.
    In the following definitions, $\min$ or $\max$ without an argument is taken with respective to $E[Y_{\cD(X)}]$
    (recall that $E[Y_{\cD(X)}]= E[E[Y_{1}|X]I\{\cD(X)=1\}+E[Y_{-1}|X]I\{\cD(X)=-1\}]$, and $E\left( Y_{-1}|X\right)$, $E\left( Y_{1}|X\right)$ satisfy
 $\cL_{-1}(X)\leq E\left( Y_{-1}|X\right) \leq \cU_{-1}(X)$, $\cL_{1}(X)\leq E\left( Y_{1}|X\right) \leq \cU_{1}(X)$, respectively), and $\cD$ belongs to the set of all deterministic rules.

 \begin{itemize}
 \item  Maximax utility (optimist): $\max_{\cD} \max E[Y_{\cD(X)}]$;

     \item  (Wald) Maximin utility (pessimist): $\max_{\cD} \min E[Y_{\cD(X)}]$;

     \item  (Savage) Minimax regret (opportunist): $\min_{\cD} \max ( E[Y_{\cD^*(X)}] - E[Y_{\cD(X)}] )$;

     \item  Hurwicz criterion: $\max_{\cD} (\alpha \max E[Y_{\cD(X)}]+ (1-\alpha) \min E[Y_{\cD(X)}])$;

     \item Healthcare decision-making: $\max_{\cD} E[E(Y_{-1}|X)+\cL(X)I\{\cD(X)=1\}]$.

 \end{itemize}
 For example, for the left panel of Figure~\ref{fig:2},
  maximax utility criterion recommends $A=1$;
 maximin utility criterion recommends $A=-1$;
 minimax regret criterion recommends $A=-1$.

Notably, all criteria in Table~\ref{table:1} reduce to $\cD^*$ under point identification. For a more complete treatment of  decision-making strategies and formal axioms of rational choice, we refer to \citet{arrow1972}. Interestingly, we note that a (deterministic) minimax regret criterion coincides with Hurwicz criterion with $\alpha=1/2$ as $\mathcal{L}\left( X\right)=\cL_1(X)-\cU_{-1}(X)$ and $\mathcal{U}\left( X\right)=\cU_1(X)-\cL_{-1}(X)$.

\begin{table}[H]
\centering
\caption{Different representations of $w(x)$ for various decision-making strategies.
 \label{table:1}}
\resizebox{1\columnwidth}{!}{
\begin{tabular}{| c| c| c|}
\hline
Decision theory criterion & $w(x)$ & The corresponding rule \\\hline
 Maximax utility (optimist) & $I(P>Q)$ &
 \parbox{3cm}{\begin{align*}
 \cD (w(x),x)&=
  \begin{cases}
                                   1 & \cU_1(x)>\cU_{-1}(x), \\
                                   -1 &  \cU_1(x)< \cU_{-1}(x). \\
  \end{cases}
   \end{align*}}
  \\\hline
 Maximin utility (pessimist) & $I(P<Q)$ &
  \parbox{3cm}{\begin{align*}
\cD (w(x),x) &=
  \begin{cases}
                                   1 & \cL_1(x)>\cL_{-1}(x), \\
                                   -1 &  \cL_1(x)< \cL_{-1}(x). \\
  \end{cases}
       \end{align*}}
 \\\hline
 Minimax regret (opportunist) & $1/2$ &
 \parbox{3cm}{\begin{align*}
 \cD (w(x),x)& =
  \begin{cases}
                                   1 & \cL(x)>0~\text{or}~\cL(x)<0< \cU(x),|\cU(x)|>|\cL(x)|, \\
                                   -1 &  \cU(x)<0~\text{or}~\cL(x)<0< \cU(x), |\cU(x)|<|\cL(x)|.
  \end{cases}
     \end{align*}}
 \\\hline
 \pbox{20cm}{ \hspace{0.7cm} Hurwicz criterion\\ ($0\leq \alpha \leq 1$: Hurwicz index)} & $\alpha I(P>Q) + (1-\alpha) I(P<Q)$ &
  \parbox{3cm}{\begin{align*}
\cD (w(x),x) &=
  \begin{cases}
                                   1 & (1-\alpha)\cL_1(x)+\alpha \cU_1(x)>(1-\alpha)\cL_{-1}(x)+\alpha\cU_{-1}(x), \\
                                   -1 &  (1-\alpha)\cL_1(x)+\alpha \cU_1(x)<(1-\alpha)\cL_{-1}(x)+\alpha\cU_{-1}(x). \\
  \end{cases}
       \end{align*}}
 \\\hline
\pbox{20cm}{Healthcare decision-making\\ ($A=-1$: standard of care)} & 0 &
 \parbox{3cm}{\begin{align*}
\cD (w(x),x) &=
  \begin{cases}
                                  1 & \cL(x)>0, \\
                                  -1 & \cL(x)<0. \\
  \end{cases}
      \end{align*}}
 \\\hline
... & ...  &... \\  \hline
Mixed strategy & a Bernoulli $w(x)$ & a stochastic $\cD(w(x),x)$ \\ \hline
\end{tabular}
}
{\scriptsize Define $P\equiv \mathcal{L}\left( x\right)
 I\left\{ \mathcal{D}(x)=1\right\} +\mathcal{L}_{-1}\left( x\right)$ and $Q \equiv  -\mathcal{U}\left( x\right) I\left\{ \mathcal{D}(x)=-1\right\} +\mathcal{L}_{1}\left( x\right)$. The arguments of $x$ and $\cD$ in $P$ and $Q$ are omitted for simplicity. To streamline the presentation, we omit the case of tiebreaking.}
\end{table}

\begin{remark}
While both
lower bound criterion and Hurwicz criterion have an index, they are conceptually and technically different.
The index $w(x)$ being a number between 0 and 1 refers to the preference of actions; with $w(x)$ being a weighted average of $I(P<Q)$ and $I(P>Q)$, the lower bound criterion balances pessimism and optimism; however, it may not be straightforward for Hurwicz criterion to balance preferences on treatments/actions.
\end{remark}

\subsection{Incorporating individualized preferences: numeric/symbolic/stochastic inputs} \label{sec:5.2}

We note that the lower bound criterion also sheds light on the process of data collection for individualized decision-making.
As individuals in the population of interest may ultimately exhibit different preferences for selecting optimal decisions, it may be unreasonable to assume that all participants share a common preference for evaluating optimality of an individualized decision rule under partial identification.
An investigator might collect participants' risk preferences over the space of rational choices to construct an individualized decision rule.
Therefore, we use the subscript $r$ (a participant's observed preference) to remind ourselves that $w_r(x)$ depends not only on $x$ but also on an individual's risk preference, that is, $r\in \mathcal{R}$ determines a specific form of $w_r(x)$ (see Table~\ref{table:1}), where $\mathcal{R}$ is a collection of different risk preferences.
Such $w_r(x)$ results in a decision rule $\cD(w_r(x),x)$ depending on both $x$ (standard individualization, e.g., in the sense of subgroup identification) and $r$ (individualized risk preferences when faced uncertainty), where $r$ can be collected from each individual.

\begin{remark}
We note that part of the elegance of this lower bound framework is that the risk preference does not come into play if there is no uncertainty about optimal decision, that is, if $0 \notin(\cL(x),\cU(x))$, regardless what $w_r(x)$ being chosen, $\cD(w_r(x),x)=\cD^*(x)$.
\end{remark}

Remarkably, the recorded index $w_r(x)$ for each $x$ could be numeric/symbolic/stochastic, that is,
fall into any of the following three categories, while the participants only need to specify a category and input a number between 0 and 1 if the first two categories are chosen:

\begin{itemize}
\item Treatment/action preferences: Input a number $\beta$ between 0 and 1 which indicates preference on treatments/actions with larger $\beta$ in favor of $A=1$. Here, $w_r(x)=\beta$.
In observational studies, most applied researchers upon observing $0\in (\cL(x),\cU(x))$ would rely on standard of care ($A=-1$) and opt to wait for more
conclusive studies, which corresponds to $\beta=0$.
In a placebo-controlled study with $A=-1$ denoting placebo, $\beta = 0$ represents concerns about safeness/aversion of treatment.

\item Utility/risk preferences: Input a number $\beta$ between 0 and 1 and let symbolic input $w_r(x)=\beta I(P>Q) + [1-\beta] I(P<Q)$,  where $\beta$ refers to the coefficient of optimism. For instance, $\beta=0$ puts the emphasis on the worst possible outcome, and refers to risk aversion; and likewise $\beta=1/2$, $1$ refer to risk neutral and risk taker, respectively.

\item  An option for opportunists who are unwilling to lose: Render $w_r(x)$ random as a Bernoulli random variable, see Section~\ref{sec:minimax_game} for details.
\end{itemize}

We highlight that the proposed index $w_r(x)$ unifies various concepts in artificial intelligence, economics, and statistics, which holds promise for providing a satisfactory regime for each individual through machine intelligence.

\subsection{A randomized minimax regret solution for opportunists}\label{sec:minimax_game}
 In this section, we consider whether an investigator/participant who happens to be an opportunist can do better in terms of protecting the worst case regret than the minimax regret approach in Table~\ref{table:1}.

An opportunist might not put all of his or her eggs in one basket.
This mixed strategy is also known as mixed portfolio in portfolio optimization.
Let $p(x)$ denote the probability of taking $A=1$ given $X=x$, by the definition of the minimax regret criterion, one essentially needs to solve the following for $p(x)$,
$$\min_{p(x)} \max([1-p(x)] \max\{\cU(x),0\} ,p(x) \max\{-\cL(x),0\}),$$
which leads to the following
solution
\begin{align*}
 p^*(x)=
  \begin{cases}
   1&  \cL(x)>0, \\
    0 &  \cU(x)<0, \\
                                  \frac{\cU(x)}{\cU(x)-\cL(x)} & \cL(x)<0<\cU(x). \\
  \end{cases}
   \end{align*}
Such a choice of $p^*(x)$ guarantees the worst case regret no more than
\begin{align*}
  \begin{cases}
    0 &  \cU(x)<0~\text{or}~\cL(x)>0, \\                                  -\frac{\cL(x)\cU(x)}{\cU(x)-\cL(x)} & \cL(x)<0<\cU(x).\\
  \end{cases}
   \end{align*}
      We formalize the above result in the following theorem.
   \begin{theorem}
   Define the stochastic policy $\widetilde \cD$ as $\widetilde \cD(x)=1$ with probability $p^*(x)$,
   the corresponding regret is bounded by
   \begin{align*}
E[Y_{\cD^*(X)}] -  E[Y_{\widetilde \cD(X)}] \leq E\left[
-\frac{\cL(X)\cU(X)}{\cU(X)-\cL(X)} I\{\cL(X)<0<\cU(X)\} \right],
   \end{align*}
   where $E[Y_{\widetilde \cD(X)}] =  E_X[ E_{\widetilde \cD} [E_{Y_{\widetilde \cD}}[Y_{\widetilde \cD(X)}|\widetilde \cD,X] |X]  ]$.
   \end{theorem}

  In contrast, by only considering deterministic rules, a minimax regret approach guarantees the worst case regret for $X=x$ which is no more than $$\min ( \max\{\cU(x),0\} ,\max\{-\cL(x),0\}).$$
  It is clear that
  \begin{align*}
   -\frac{\cL(x)\cU(x)}{\cU(x)-\cL(x)} < \min\{-\cL(x),\cU(x)\} ~~~~\text{if}~~~~ \cL(x)<0<\cU(x).
  \end{align*}
  Therefore, the proposed mixed strategy gives a sharper minimax regret bound than \citet{zp} and \citet{pz}, and therefore is sharper than any deterministic rules.

\begin{remark}
The result in this section does not necessarily rely on $\cL(x)$ being
defined as $\cL_1(x) - \cU_{-1}(x)$ and $\cU(x)$ being defined as $\cU_1(x) - \cL_{-1}(x)$.
\end{remark}

\begin{remark}
The proposed mixed strategy leads to $w(x)$ or $w_r(x)$ a  Bernoulli random variable with probability $p^*(x)$, and therefore a stochastic rule $\cD(w(x),x)$ or $\cD(w_r(x),x)$ assigning 1 with probability $p^*(x)$. Note that
$w_r(x)$ being a
Bernoulli random variable with parameter $p(x)$, and $w_r(x)$
being a scalar $p(x)$ are fundamentally different: The former one provides a stochastic decision rule. In other words, participants with the same $x$ can receive different recommendations; while the latter one leads to a deterministic rule. That is, all participants with the same $x$ receive the same recommendation.
\end{remark}

\subsection{No universal optimality for decision-making under partial identification} \label{sec:no_optimality}

As can be easily seen from Table~\ref{table:1} as well as Section~\ref{sec:minimax_game}, there is a mismatch between deterministic/randomized minimax regret and maximin utility. In fact, each of the three rules corresponds to a different decision strategy.
Such mismatch is a distinctive feature of partial identification.

On the one hand, it is notable that $\{\cL(x),\cU(x)\}$ provides complementary information to the analyst as it might inform the analyst as to when he/she might refrain from making a decision; mainly, if such an interval includes zero so that there is no evidence in the data as to whether the action/treatment is on average beneficial or harmful for individuals with that value of $x$.
One might need to conduct randomized experiments in order to draw a causal conclusion if $0\in (\cL(x),\cU(x))$.
On the other hand, the decision-making must in general be considered a game of four numbers $\{\cL_1(x),\cL_{-1}(x),\cL(x), \cU(x) \}$
rather than two, for example, $\{\cL_1(x),\cL_{-1}(x)\}$ or $\{\cL(x),\cU(x)\}$.

From the above point of view,
the concept of optimality of a decision rule under partial identification cannot be absolute, rather, it is relative to a particular choice of decision-making criterion, whether it is minimax, maximax, maximin, and so on.
Furthermore, an individualized decision rule might incorporate participants' risk preferences as it might be unreasonable to assume everyone shares a common preference.
 In the Appendix, we provide expressions for the minimum utility, maximum regret, and maximum misclassification rate of certain `optimal' rules in Table~\ref{table:1} (including maximin utility and deterministic/randomized minimax regret rules) for practical uses.

\section{A paradox: 1+1<2}\label{sec:6}

In this section, we provide an interesting paradox regarding the use of partial identification to conduct individualized decision-making.
To streamline our presentation, we use (deterministic) minimax regret rule as a running example, however, any rule $\cD\in \cD^{\text{opt}}$ can suffer the same paradox. To simplify exposition, we consider the case with no $U$, that unbeknownst to the analyst, unmeasured confounding is absent.
We consider the following model with covariate $X$ (e.g., female/male) distributed on $\{0, 1\}$ with equal probabilities,
\begin{align*}
\Pr(Y=1|X,A) &= X/16 + 1/5A + 1/15,\\
\Pr(A=1|X,Z) &= X/16 + 2/5Z + 1/2,\\
Z & \sim \text{Bernoulli}(1/2).
\end{align*}
With a slight abuse of notation, we use $0,1$ coding for $Z,A$ here.
It is easy to see that the optimal rule is $\cD^*=1$ for the entire population. After a simple calculation,
the Balke-Pearl conditional average treatment effect bounds for $X=0,1$ both contain zero with
$|\cL(0)|<|\cU(0)|$ and $|\cL(1)|>|\cU(1)|$.
The Balke-Pearl average treatment effect bounds marginalizing over $X$ also contain zero and $|\cL|<|\cU|$.

As it is unbeknownst to the analyst whether unmeasured confounding is present or whether $X$ is an effect modifier, there are several possible strategies for analyzing the data.

\begin{enumerate}
\item  If one is concerned about individualized decision-making but does not worry about unmeasured confounding, one runs a standard regression type analysis and gets the right answer.

\item If one is concerned about unmeasured confounding but is only interested in decision-making based on the population level (i.e., based on average treatment effect analysis), one can obtain IV bounds on the average treatment effect and also get the right answer.

\item  If one is concerned about individualized decision-making and also worries about unmeasured confounding, one gets the wrong answer for a subgroup.
\end{enumerate}

We summarize results of the above strategies of analyses in  Table~\ref{table:2}.

\begin{table}[H]
\begin{tabular}{lll}
    & $X=0$  & $X=1$    \\
(1) & $\surd$ & $\surd$ \\
(2) & $\surd$ & $\surd$ \\
(3) & $\surd$ & $\times$
\end{tabular}
\caption{Correct/incorrect decisions using three types of data analyses. \label{table:2}}
\end{table}

As can be seen from the table, mixing up two very difficult domains (individualized recommendation + unmeasured confounding) might make life harder (1 + 1 < 2).
There are several lessons one can learn from this paradox:

a) A comparison between (1) and (3): It would be a good idea to first conduct a standard analysis (e.g., assume Assumption~\ref{asm:unconfoundedness}) or other point identification approaches \citep[e.g., assume Assumption~7 of][]{cui2020} and then use IV bounds as a sanity check or say policy improvement;

b) A comparison between (2) and (3): The paradox sheds light on the clear need for carefully distinguish variables
used to make individualized decisions from variables used to address confounding concerns;  similar to but different from Simpson's paradox, the aggregated and disaggregated answers can be opposite for a substantial subgroup.

c) (3) by itself: It might be a rather risky undertaking to narrow down an interval estimate to a definite decision given the overwhelming uncertainty; overly accounting for unmeasured confounding might erroneously recommend a sub-optimal decision to a subgroup.

As motivated by the comparison between (1) and (3),
we formalize the policy improvement idea following \citet{kallus2018confounding}. Note that minimizing the worst-case possible regret against a baseline policy $\cD_0$ would improve upon those individuals for whom $\cD_0(X)=-1, \cL(X)>0$ and $\cD_0(X)=1, \cU(X)<0$.
We revisit the real data example in Section~\ref{sec:real}. We first run a standard analysis (random forest: $Y$ on $X,A$) and obtain $\cD_0(X)=\sign\{\Pr(Y|X,A=1)-\Pr(Y|X,A=-1)\}$; among 3,010 subjects, 2,106 have $\cD_0(X)=1$ and 904 have $\cD_0(X)=-1$.
Then we calculate IV conditional average treatment effect bounds, and there are 323 subjects with $\cL(X)>0$ and 45 subjects with $\cU(X)<0$.
Then we use IV bounds as a sanity check/improvement: Only $4$ subjects with $\cD_0(X)=-1$ switch to $1$,  and $8$ subjects with $\cD_0(X)=1$ switch to $-1$. Therefore, for most subjects in this application, the IV bounds do not necessarily invalidate the standard regression analysis, while IV bounds are still helpful to validate/invalidate decisions for a subgroup.

\section{Discussion}
In this article, we illustrated how one might pursue individualized decision-making using partial identification in a comprehensive manner.
We established a formal link between individualized decision-making under partial identification and classical decision theory by considering a lower bound perspective of value/utility function. Building on this unified framework, we provided a novel minimax solution for opportunists who are unwilling to lose.  We also pointed out that there is a mismatch between maximin utility and minimax regret.
Moreover, we provided an interesting paradox to ground several interesting ideas on individualized decision-making and unmeasured confounding.  To conclude, we list the following points that might be worth considering in future research.

\begin{itemize}
 \setlength\itemsep{0.4em}

  \item As the proper use of multiple IVs is of growing interest in a lot of applications including
statistical genetics studies, one could possibly construct multiple IVs and then try to find multiple bounds to conduct a better sanity check or improvement.
Another possibility is to strengthen multiple IVs \citep{zubizarreta2013stronger,ertefaie2018quantitative}. A stronger IV might provide a tighter bound, and therefore a sign identification may be achieved \citep{cui2020necessary}.

    \item Including additional covariates
  which are associated with $A$ or $Y$ for stratification and then marginalizing over these covariates would potentially give a tighter bound.
   Therefore, carefully choosing variables used to stratify (which can be the same as decision variables or a larger set of variables) might be of interest for both theoretical and practical purposes.

  \item The proposed minimax regret method by leveraging a randomization scheme and other strategies in Table~\ref{table:1} might be of interest in optimal control settings such as reinforcement learning and contextual bandit where exploitation and exploration are under consideration.
  In addition, given observational data in which a potential IV is available, one can use different strategies to construct an initial randomized policy for use in a reinforcement learning and bandit algorithm.

  \item One important difference between decision-making with IV partial identification and classical decision theory is the source of uncertainty. For the former one, unmeasured confounding creates uncertainty, and overthinking confounding might create overwhelming uncertainty.
  Therefore, to better assess the uncertainty, it would also be of great interest to formalize a sensitivity analysis procedure for point identification such as under assumptions of no unmeasured confounding or no unmeasured common effect modifiers \citep{cui2020}. A similar question has also been raised by \citet{han2020}.

 \end{itemize}

\subsection*{Disclosure Statement}
The author is supported by NUS grant R-155-000-229-133.

\subsection*{Acknowledgments}
The
author is thankful to three referees, associate editor, and Editor-in-Chief for
useful comments, which led to an improved manuscript.



\appendix
\numberwithin{equation}{section}

\section{Derivation of lower bounds of value function}
The following was originally derived in \citet{cui2020}. It is helpful to provide it here.

\begin{proof}
Note that
\begin{align*}
E\left[ Y_{\mathcal{D}(X)}|X\right] &=E\left( Y_{1}|X\right) I\left\{
\mathcal{D}(X)=1\right\} +E\left( Y_{-1}|X\right)I\left\{
\mathcal{D}(X)=-1\right\},\\
E\left[ Y_{\mathcal{D}(X)}|X\right] &=E\left( Y_{1}-Y_{-1}|X\right) I\left\{
\mathcal{D}(X)=1\right\} +E\left( Y_{-1}|X\right),\\
E\left[ Y_{\mathcal{D}(X)}|X\right] &=E\left( Y_{-1}-Y_{1}|X\right) I\left\{
\mathcal{D}(X)=-1\right\} +E\left( Y_{1}|X\right).
\end{align*}
By  $\cL_{-1}(X)\leq E\left( Y_{-1}|X\right) \leq \cU_{-1}(X)$ and  $\cL_{1}(X)\leq E\left( Y_{1}|X\right) \leq \cU_{1}(X)$, one has the following bounds,
\begin{align}
&(1-w(X)) [\mathcal{L}\left( X\right)
 I\left\{ \mathcal{D}(X)=1\right\} +\mathcal{L}_{-1}\left( X\right)]
+ w(X) [ -\mathcal{U}\left( X\right) I\left\{ \mathcal{D}(X)=-1\right\} +\mathcal{L}_{1}\left( X\right)] \nonumber \\
&\leq \mathcal{L}_{1}(X) I\{\cD(X)=1\} +\mathcal{L}_{-1}(X) I\{\cD(X)=-1\} \leq E\left[ Y_{\mathcal{D}
(X)}|X\right], \label{s:1}
\end{align}
where $0 \leq w(x)\leq 1$ for any $x$.
Therefore, we complete the proof by taking expectations on both sides of Equation~\eqref{s:1}.
\end{proof}

\section{Minimum utility, maximum regret,  and  maximum misclassification rate of several `optimal' rules}

We give the minimum value function, maximum regret, and maximum misclassification rate over $\cD\in \cD^{opt}$ expressed in terms of the observed data:
\begin{align*}
& E[\max(\cL_{-1}(X),\cL_{1}(X)) I\{0\notin (\cL(X),\cU(X))\}+\min(\cL_{-1}(X),\cL_{1}(X)) I\{0\in (\cL(X),\cU(X))\}],\\
& E[\max(|\cL(X)|,|\cU(X)|) I\{0\in (\cL(X),\cU(X))\}],\\
& E[I\{0\in (\cL(X),\cU(X))\}],
\end{align*}
respectively. While the maximum misclassification rate remains the same, the minimum value function and maximum regret for a given $\cD$ can be different. For instance, the minimum value function and maximum regret of the maximin rule in Table~\ref{table:1} are:
\begin{align*}
& E[\max(\cL_{-1}(X),\cL_{1}(X))],\\
&
 E\Big[ \big[|\cL(X)|I\{\cL_{-1}(X)<\cL_{1}(X)\} +
|\cU(X)|I\{\cL_{-1}(X)>\cL_{1}(X)\}\big]I\{0\in (\cL(X),\cU(X))\}\Big],
\end{align*}
respectively. The minimum value function and maximum regret of the minimax rule in Table~\ref{table:1} are:
\begin{align*}
& E\Big[\max(\cL_{-1}(X),\cL_{1}(X)) I\{0\notin (\cL(X),\cU(X))\}\\
& ~~~~ +\big[\cL_{1}(X) I\{|\cL(X)|<|\cU(X)|\} + \cL_{-1}(X)I\{|\cL(X)|>|\cU(X)|\}  \big]I\{0\in (\cL(X),\cU(X))\}\Big],\\
& E[\min(|\cL(X)|,|\cU(X)|) I\{0\in (\cL(X),\cU(X))\}],
\end{align*}
respectively. The minimum value function and maximum regret of the randomized minimax rule in Section~\ref{sec:minimax_game} are:
\begin{align*}
& E\bigg[\max(\cL_{-1}(X),\cL_{1}(X)) I\{0\notin (\cL(X),\cU(X))\}\\
& ~~~~ +\left[\cL_{1}(X) \frac{\cU(X)}{\cU(X)-\cL(X)} + \cL_{-1}(X)\frac{-\cL(X)}{\cU(X)-\cL(X)}  \right]I\{0\in (\cL(X),\cU(X))\}\bigg],\\
& E\left[ -\frac{\cL(X)\cU(X)}{\cU(X)-\cL(X)} I\{0\in (\cL(X),\cU(X))\}\right],
\end{align*}
respectively.

\printbibliography

\end{document}